\documentclass[usenatbib]{mnras}
\usepackage{amsmath}
\usepackage{amssymb}
\usepackage{color}
\usepackage{epsfig}
\usepackage{graphicx}
\usepackage{url}
\usepackage{hyperref}

\newcommand{\xhi}{x_{\rm HI}}
\newcommand{\xhii}{x_{\rm HII}}
\newcommand{\gammahi}{\Gamma_{\rm HI}}
\newcommand{\lambdanu}{\lambda_{\nu_0}}
\newcommand{\hstar}{\mathcal{H}_*}

\title[Do CRs heat the early IGM?]{Do cosmic rays heat the early intergalactic medium?}

\author[N.~Leite et al.]{N. Leite,$^{1}$
C. Evoli,$^{2}$ 
M. D'Angelo,$^{2}$ 
B. Ciardi,$^{3}$ 
G. Sigl,$^{1}$ 
A. Ferrara$^{4}$ \\
$^{1}$~II. Institute for Theoretical Physics, University of Hamburg, Luruper Chaussee 149, 22761 Hamburg, Germany\\
$^{2}$~Gran Sasso Science Institute (GSSI), Viale Francesco Crispi 7, L'Aquila 67100, Italy\\
$^{3}$~Max-Planck-Institut f\"ur Astrophysik, Karl-Schwarzschild-Stra{\ss}e 1, D-85748 Garching bei M\"unchen, Germany\\
$^{4}$~Scuola Normale Superiore, Piazza dei Cavalieri 7, I-56126 Pisa, Italy\\
}

\begin{document}

\date{}

\pagerange{\pageref{firstpage}--\pageref{lastpage}} \pubyear{2017}

\maketitle

\label{firstpage}

\begin{abstract}
Cosmic rays (CRs) govern the energetics of present-day galaxies and might have also played a pivotal role during the Epoch of Reionization. In particular, energy deposition by low-energy ($E \lesssim 10$ MeV) CRs accelerated by the first supernovae, might have heated and ionized the neutral intergalactic medium (IGM) well before ($z \approx 20$) it was reionized, significantly adding to the similar effect by X-rays or dark matter annihilations.
Using a simple, but physically motivated reionization model, and a thorough implementation of CR energy losses, we show that CRs contribute negligibly to IGM ionization, but heat it substantially, raising its temperature by $\Delta T=10-200$ K by $z=10$, depending on the CR injection spectrum. Whether this IGM pre-heating is uniform or clustered around the first galaxies depends on CR diffusion, in turn governed by the efficiency of self-confinement due to plasma streaming instabilities that we discuss in detail. This aspect is crucial to interpret future HI 21 cm observations which can be used to gain unique information on the strength and structure of early intergalactic magnetic fields, and the efficiency of CR acceleration by the first supernovae.     
  
\end{abstract}

\section{Introduction}\label{sec:intro}

Sometimes during the first billion cosmic years the first stars formed. Ionizing photons emitted by these sources induced a major phase transition from the otherwise cold and neutral state in which the intergalactic medium (IGM) was left after recombination ($z\lesssim 500$) to the warm/ionized state we measure today. The transition phase is known as the Epoch of Reionization (EoR).
Understanding the detailed EoR physics is one of the primary goals of present-day cosmology~\citep{2001PhR...349..125B,2013ASSL..396...45Z,2014arXiv1409.4946F,2016ASSL..423.....M}. 

To date, we only have a broad-brush picture of such important event. Nevertheless sizable steps forward have been made in the last decade. 
High redshift ($z \lesssim 6$) quasar absorption line experiments, probing the physical state of the neutral IGM, indicate that reionization was complete by $z \sim 5.7$~\citep{2006AJ....131.1203F}.
The CMB polarization measurements made by PLANCK~\citep{2016A&A...594A..13P} have set tight bounds to the value of the free electron scattering optical depth, $\tau_{\rm e} = 0.066 \pm 0.016$. Translating this value into a reionization redshift, $z_r$, yields uncertain results as the conversion depends from the (yet unknown) reionization history. In the popular instantaneous reionization model it is $7.8 < z_r < 8.8$, with an upper limit on the EoR duration of $\Delta z < 2.8$. Thus, reionization might have been a remarkably fast ($\le400$ Myr) process.
  
Great advances in the understanding of the EoR history and physics are expected from a number of upcoming observations of the redshifted HI 21-cm line signal from these epochs \citep{2003ApJ...596....1C,2006PhR...433..181F,2010ARA&A..48..127M,2012RPPh...75h6901P}. 

Several experiments are attempting to measure the 21-cm signal from the EoR using low-frequency radio interferometers. These include GMRT\footnote{\url{http://gmrt.ncra.tifr.res.in}}, LOFAR\footnote{\url{http://www.lofar.org}}, MWA\footnote{\url{http://www.mwatelescope.org}}, PAPER\footnote{\url{http://eor.berkeley.edu}}, HERA\footnote{\url{http://reionization.org}} and, in the future, SKA\footnote{\url{http://www.skatelescope.org}}.
Measuring such signal will provide direct information on the pre-reionization ages. The 21-cm signal will inform us of the HI spin temperature as a function of redshift. 
As such it will probe the efficiency of gas heating and ionization mechanisms that might have played a role from recombination to the time at which the first light sources appeared (the so-called "Dark Ages"). 

The first stars and galaxies are likely to be the main sources of the ionizing UV ~\citep{2005SSRv..116..625C}. 
Both theoretical arguments and numerical studies suggest that the first generation stars (known as Pop~III stars) were more massive than present day ones and metal-free. These short-lived stars polluted the surrounding gas with metals, inducing a rapid transition to a cosmological star formation rate (SFR) dominated by the present-day, Pop~II/I, stars~\citep{2002ApJ...571...30S,2006MNRAS.369.1437S}.

However, sources of higher energy X-ray photons might also have been present. X-rays have a far longer mean free path than lower energy UV photons. Therefore, these photons are able to travel significantly larger distances in the neutral IGM and release a significant amount of energy, eventually increasing the temperature of the intergalactic gas. 
Potential X-ray sources are quasars, supernovae and X-ray binaries. However, very little is known about their abundances, evolution and spectra, especially at these high redshifts~\citep{2012MNRAS.426.1349M}.

An additional contribution to IGM heating has been proposed by~\citet{2012ApJ...752...23C,2016ApJ...833..118C}. 
The deposition of kinetic energy into the IGM via plasma instabilities~\citep{2012ApJ...758..102S} triggered by TeV photons from blazars may yield a higher heating rate than photoheating for $z\lesssim 6$.
Such an effect is more relevant in the intergalactic voids, i.e. in less dense regions, of the ionized medium and allows to better reproduce Lyman-$\alpha$ forest observables over the redshift range $2 < z < 3$~\citep{2012MNRAS.423..149P}.
On the other hand, recent numerical simulations have been suggesting that this mechanism could be fairly ineffective in the IGM conditions~\citep{2014ApJ...787...49S}, although we note that 
there is still no general consensus on this.

Similarly to X-ray photons, cosmic rays (CRs) -- accelerated from the shocks created by the exploding supernovae and promptly released in the IGM -- could also efficiently deposit their energy in the form of gas heating~\citep{1993MNRAS.265..241N, 2015MNRAS.454.3464S}. 
Curiously, such heating source has received relatively little attention in the literature so far, in spite of the fact that it is known since decades that in local galaxies low-energy CRs play a key overall role in the energetics of these systems and, in particular, in regulating the ionization and thermal state of the interstellar gas. 
 
At higher redshifts, CRs should be able to escape their host galaxies by advection or by diffusion before they lose a significant fraction of their energy within the halo~\citep{2008ApJ...673..676R}. 
As they propagate in the IGM, CRs interact with the surrounding environments mainly via H/He photoionization and Coulomb collisions with free electrons, and by doing so they deposit thermal energy in the gas.
Both mechanisms imply that CRs could contribute to the thermal history of the IGM~\citep{2005ICRC....9..215S}. 

Earlier studies of the impact of CRs on the high-redshift IGM mainly focused on the cosmological CRs originated by Pop~III stars~\citep[e.g.,][]{2006ApJ...651..658R}. 
Recently, the authors of~\citet{2015MNRAS.454.3464S} found that low-energy particles ($E \lesssim 30$ MeV) are capable of increasing the IGM temperature to 10-100 K before standard heating sources, such as galaxies and quasars, appear.

In the present work, we reanalyze the role played by CRs from high-$z$ Pop~II stars.
In fact, it has been shown that in realistic models of galaxy formation, chemical feedback suppresses metal-free star formation in the self-enriched progenitors, and although Pop III star formation can in principle persist down to $z \sim 3 - 4$, Pop II stars dominate the SFR at any redshift~\citep{2007MNRAS.382..945T,2014MNRAS.440.2498P}. 
Moreover, Pop~II stars are observed in the local Universe and their properties can be much more robustly constrained.

In \S\ref{sec:galaxies}, we build a simple reionization model and link it to the cosmic star formation rate. In \S\ref{sec:cr} we introduce the treatment of CR production, energy losses and propagation. With these ingredients, in \S\ref{sec:results} we compute the CR contribution to IGM heating, and explore the implications of the spatial dependence of the temperature increment. Results and assumptions are discussed in \S\ref{sec:conclusions}. We use the cosmological parameters: $h = 0.678$, $\Omega_m = 0.308$, $\Omega_\Lambda = 0.692$, $\Omega_b h^2 = 0.0223$, $n_s = 0.968$ and $\tau_e = 0.066$~\citep{2016A&A...594A..13P}.

\section{Reionization by early galaxies}\label{sec:galaxies}

In this Section, we assume that galaxies are the primary reionization sources. This scenario is supported by observations of high-redshift galaxies, which would be able to reionize the Universe by $z = 6$, provided that a substantial fraction of their ionizing emission escapes into the IGM~\citep{2010Natur.468...49R}.
  
Our model implements the basic physical processes required to properly model IGM evolution in the presence of the ionizing radiation from galaxies. Such formalism allows us to track the IGM thermal and ionization history, and to reproduce the available observational EoR data. A more detailed treatment can be obtained by means of numerical simulations, e.g. in~\citet{2007MNRAS.376..534I,2012MNRAS.423..558C,2016ApJ...821...50G}. 

\subsection{Star Formation Rate} \label{sec:SFR}

We assume that the star formation rate per unit of stellar mass and comoving volume (SFR) inside a DM halo is proportional to its mass, $\dot{\rho}_* (M_h) \propto M_h$, and it occurs on a free-fall time-scale, 
\begin{equation}
t_\mathrm{ff} = \sqrt{\frac{3\pi}{32 G_N \rho_m}} \, , 
\end{equation}
where $G_N$ is the gravitational constant 
and $\rho_m$ is the average mass density inside the virial radius of the halo %$r_{\rm vir}$, 
\citep{1998ApJ...495...80B}.
We further assume that stars only form in Ly$\alpha$ cooling halos \citep[][]{2000ApJ...539...20B}, i.e. those above a mass   
\begin{equation}
M_{\rm Ly\alpha}(z) \sim 10^8 M_\odot \left( \frac{10}{1+z} \right)^{3/2} \, .
\end{equation}

Radiative feedback is expected to quench SF in halos with circular velocity at the virial radius, $V_c$ smaller than a critical value $\bar{V}_c$. The corresponding minimum mass, $M_{\rm rf}(z)$, is obtained according to the relationship~\citep{2001ApJ...555...92M}
\begin{equation}
\bar{V}_c = 24 \, \text{km/s} \left( \frac{M_{\rm rf}}{10^8 \, \text{M}_\odot} \right)^{1/3} \left(\frac{1+z}{10}\right)^{1/2}\, .
\end{equation}

We can then write the cosmological SFR per unit \emph{comoving} volume at redshift $z$ as
\begin{equation} \label{eq:SFRz}
\dot{\rho}_* (z) = f_* \frac{\Omega_b}{\Omega_{m}} \int_{M_\mathrm{ min}(z)}^{\infty} dM_h \,\frac{M_h}{t_\mathrm{ ff}(M_h)} \, \frac{dN}{dM_h \, dV} \, ,
\end{equation}
where $f_*$ is the star-forming efficiency, $M_{\rm min}$ is the maximum between 
$M_{\rm Ly\alpha}$ and $M_{\rm rf}$. $\Omega_b$ and $\Omega_m$ are the density parameters of baryonic and total matter, respectively, and $dN/(dM_h dV)$ gives the number density of haloes within a mass range ($M_h, M_h + dM_h$).
To estimate the DM halo mass function we use the Press-Schechter formalism augmented by the Sheth-Tormen correction for ellipsoidal collapse~\citep{2001MNRAS.323....1S}.

\begin{figure}
\begin{center}
\includegraphics[width=.99\columnwidth]{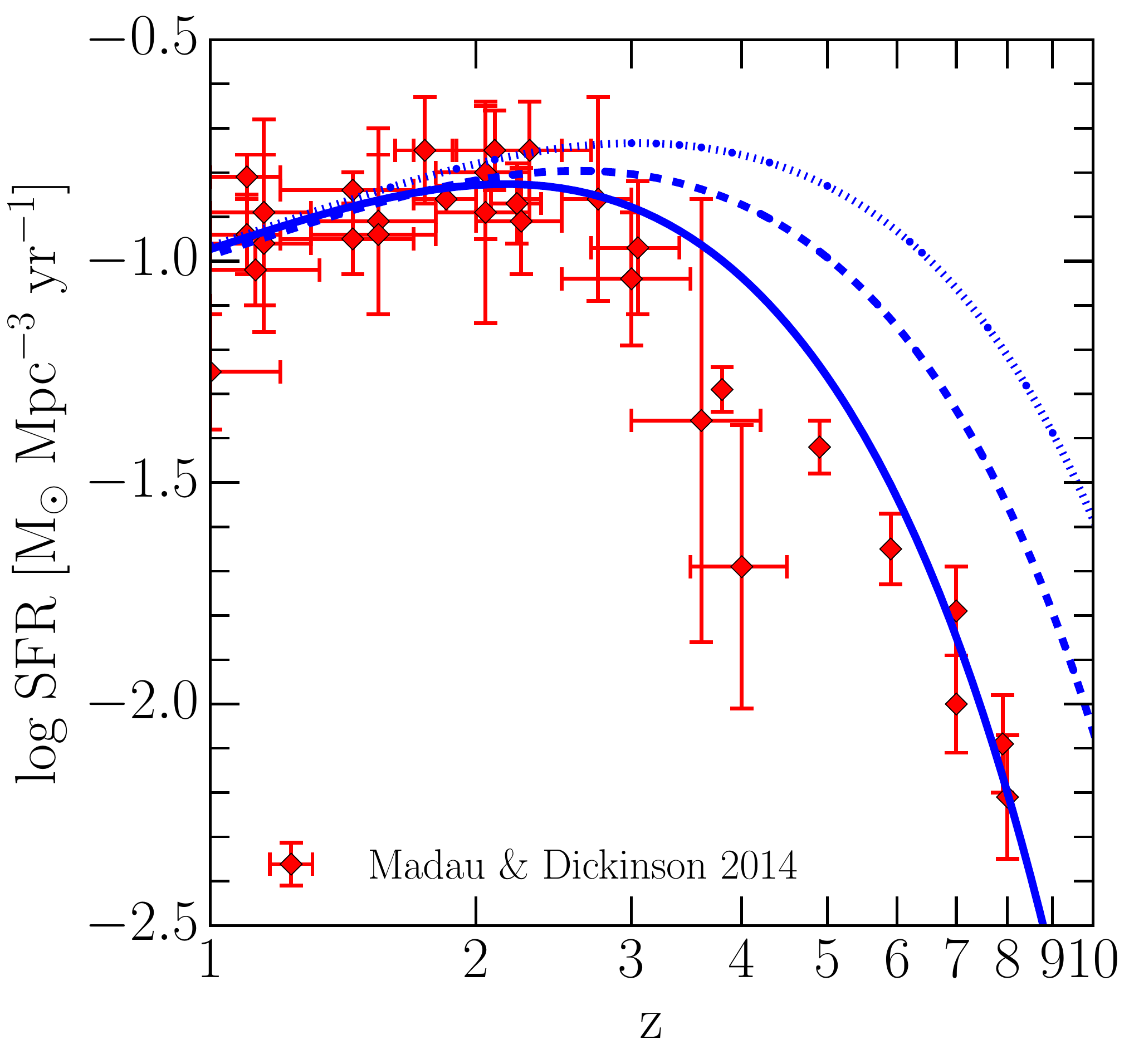}
\caption{The comoving SFR predicted by Eq.~\eqref{eq:SFRz} for different choices of the model parameters and compared with the SFR measurements by~\citet{2014ARA&A..52..415M}: $f_* = 0.02$ and $\bar{V}_c = 100$~km/s (solid curve), $f_* = 0.012$ and $\bar{V}_c = 50$~km/s (dashed curve), $f_* = 0.01$ and $\bar{V}_c = 30$~km/s (dotted curve).}
\label{fig:SFR}
\end{center}
\end{figure}

As we show in Fig.~\ref{fig:SFR}, a value of $f_* = 0.02$ and $\bar{V}_c = 100$~km/s allows us to reproduce the SFR measurements reported by~\citet{2014ARA&A..52..415M}.

\subsection{Reionization history} \label{sec:reion}

We denote the neutral hydrogen (HI) fraction by $x_{\rm H \small{I}} = n_{\rm H \small{I}}/n_{\rm H}$ and the ionized fraction by $x_{\rm H \small{II}} = n_{\rm H \small{II}}/n_{\rm H}$ with $x_{\rm H \small{I}} + x_{\rm H \small{II}} = 1$ where $n_{x}$ is the number density of species $x$.

It is further assumed that the ionization fraction of singly ionized helium and hydrogen are equal,
$x_{\rm He \small{II}} = x_{\rm H \small{II}}$.

We can then write the evolution equation for the ionization fraction in terms of the photo-ionization ($\gammahi$) and the recombination ($R$) rates as
\begin{equation} \label{eq:ion_frac}
\frac{d \xhii}{dz} = \frac{dt}{dz} \left[ \xhi \gammahi - R \right].
\end{equation} 
Since stellar radiation contributes to ionize the IGM with photons of energy higher than the ionization threshold $I_H = h \nu_0 = 13.6$ eV, one can write the ionization rate as~\citep{2005MNRAS.361..577C}
\begin{equation} \label{eq:ion_rate}
\gammahi (z) = \int_{\nu_0}^\infty d\nu \, \lambda_{\rm HI}(z,\nu) \sigma_{\rm HI}(\nu) \dot{n}_{\gamma}(z, \nu) ,
\end{equation}
where the ionization cross-section is $\sigma_{\rm HI}(\nu) =  \sigma_0 (\nu/\nu_0)^{-3}$, with $\sigma_0 = 6.3 \times 10^{-18}$ cm$^2$.

\begin{figure}
\begin{center}
\includegraphics[width=.99\columnwidth]{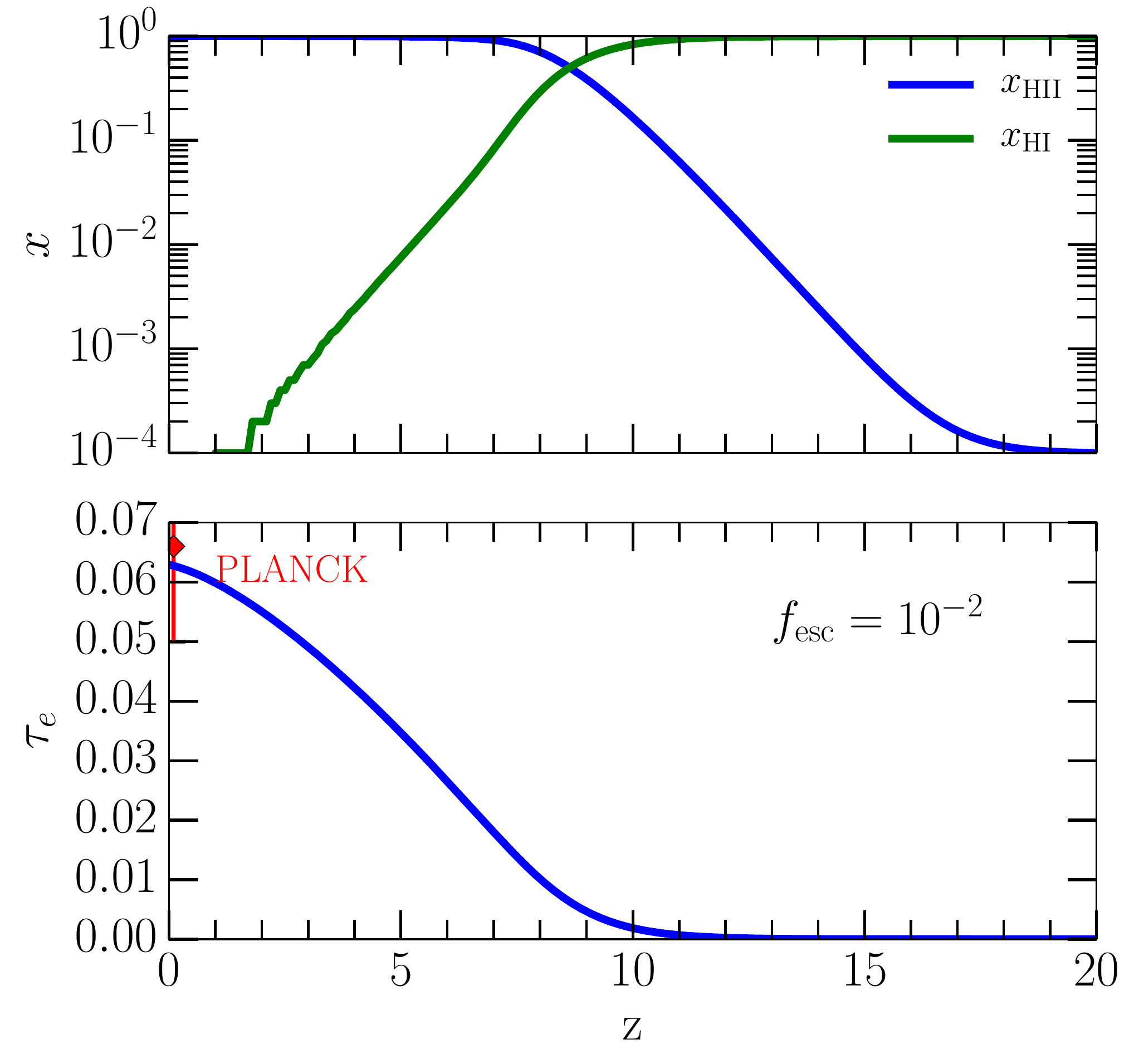} 
\caption{The IGM ionization fraction as a function of redshift for $f_{\rm esc} = 10^{-2}$ (top panel). Total optical depth computed from Eq.~\eqref{eq:tau_e} compared with the PLANCK measurement~\citep{2016A&A...594A..13P} (bottom panel).}
\label{fig:optdepth}
\end{center}
\end{figure}

The mean free path of hydrogen ionizing photons depends on the distribution of absorbing Lyman limit systems, which can be computed through the distribution of the column density $N_{H \small{I}}$ \citep{1993MNRAS.262..499P}. Assuming the distribution of absorbers to be given by $f(N_{H\small{I}})dN_{H\small{I}} \propto N_{H\small{I}}^{-3/2}dN_{H\small{I}}$ \citep{2008ApJ...688...85F}, the mean free path increases with frequency as in \citet{2003ApJ...597...66M},
\begin{equation}
\lambda_{\rm HI}(z,\nu)=\lambdanu (z) \left( \frac{\nu}{\nu_0} \right)^{3/2},
\end{equation}
with $\lambdanu (z)$ depending on the size and topology of the ionized regions. 

In~\citet{2003ApJ...584..110S}, 
%\textrm{\red Outdated. Please check recent works by Bolton and/or Haehnelt + Sobacchi} 
$\lambda_{\nu_0}(z)$ is derived from the column density distribution of the Lyman-limit systems and is found to be a rapidly evolving function of $z$. We then assume 
\begin{equation}\label{eq:lambdanu}
\lambdanu (z) \approx 39 \left( \frac{1+z}{4} \right)^{-5} \, {\rm Mpc} \, .
\end{equation}

Finally in Eq.~\eqref{eq:ion_rate}, the proper specific density rate of ionizing photons production is given by
\begin{equation}
\dot{n}_{\gamma}(z, \nu) = f_{\rm esc} \frac{dN_\gamma}{dM d\nu}  \, \dot{\rho}_*(z) (1+z)^3 \, ;
\end{equation}
$f_{\rm esc}$ is the escape fraction of ionizing photons from galaxies, and $dN_\gamma/(dM d\nu)$ the specific number of photons produced per unit mass of Pop II stars formed. 
Here we assume that the UV stellar spectrum is a power law $\propto \nu^{-\beta}$ and that, integrating over frequency, $dN_\gamma/dM = 8.05\times 10^{60} M_\odot^{-1}$~\citep{2005MNRAS.361..577C}. 

Putting all together, Eq.~\eqref{eq:ion_rate} can then be simplified to
\begin{eqnarray} \label{eq:ion_ratez}
\gammahi (z) & = & \left(\frac{\beta - 1}{\beta + 1/2}\right) \sigma_{0} f_{\rm esc} \frac{dN_\gamma}{dM} \lambdanu (z) \dot{\rho}_*(z) (1+z)^3 \\
& \approx & 5 \times 10^{-8} \, {\rm s}^{-1} 
\left(\frac{f_{\rm esc}}{10^{-2}} \right) 
\left( \frac{\dot{\rho}_*(z)}{M_\odot \, {\rm Mpc}^{-3} \, {\rm yr}^{-1}} \right) 
\left( 1 + z \right)^{-2} \, , \nonumber
\end{eqnarray}
where $\beta = 5$ is a typical value for Pop II stars. 

The recombination rate can be expressed as %\cite{2011MNRAS.411..955M}
\begin{eqnarray} \label{eq:rec_ratez}
R(z) & = & \alpha_A (T_k^{\rm i}) C n_e (z) \xhii (z) \nonumber\\ 
     & = & \alpha_A(T_k^{\rm i}) C (1 + \chi_{\rm He}) n_{\rm H} x_{\rm H \small{II}}^2 \, , 
\end{eqnarray}
where $\alpha_A$
is the case-A recombination coefficient, $n_e(z) = \xhii (z) (1+\chi_{\rm He}) n_{\rm H} (z)$ is the total electron number density 
and $\chi_{\rm He}$ the cosmic helium fraction (in density).
For the clumping factor $C\equiv \langle n_{\rm H \small{II}}^2 \rangle / \langle n_{\rm H \small{II}} \rangle ^2 \approx 2$ we take the fiducial average given by~\citet{2011MNRAS.410.1703F}.

The evolution of the gas temperature in the ionized regions ($T_k^{\rm i}$) is given by the combination of all the cooling and heating processes~\citep[see, e.g.,][]{2001MNRAS.327L...5N}
\begin{equation}\label{eq:kintemp}
\frac{dT_k^{\rm i}}{dz}
= \left. \frac{dT_k^{\rm i}}{dz} \right|_{\rm ex} 
+ \left. \frac{dT_k^{\rm i}}{dz} \right|_{\rm ion} 
+ \left. \frac{dT_k^{\rm i}}{dz} \right|_{\rm ph} \, ,
\end{equation}
where the cooling due to the expansion of the universe can be written as
\begin{equation}
\left. \frac{dT_k^{\rm i}}{dz} \right|_{\rm ex} 
= \frac{2T_k^{\rm i}}{1+z} \, ,
\end{equation}
and the heating due to the change in the internal energy, corresponding to the change in the total number of gas particles due to ionizations, where He ionizations are assumed to be negligible, is
\begin{equation}
\left. \frac{dT_k^{\rm i}}{dz} \right|_{\rm ion} 
= - \frac{T_k^{\rm i}}{1+x_e}\frac{dx_e}{dz} \, ,
\end{equation}
with $x_e=n_e/n_H$.
The last term is the heat gain by the gas particles from the surrounding radiation field
\begin{equation} \label{eq:kintemp}
\left. \frac{dT_k^{\rm i}}{dz} \right|_{\rm ph} 
= \frac{2}{3 k_B (1+x_e)} \frac{dt}{dz} \hstar \, ,
\end{equation}
where $k_B$ is the Boltzmann constant and $\hstar$ is the photoheating rate per baryon.

Analogously to Eq.~\eqref{eq:ion_rate}, the latter can be written as
\begin{equation}
\hstar (z) =  \xhi (z) \int_{\nu_0}^\infty d\nu \, \lambda_{\rm HI}(z,\nu) \sigma_{\rm HI}(\nu) \dot{n}_{\gamma}(z, \nu)  h \left(\nu - \nu_0 \right) ,
\end{equation}
which we can integrate over frequency, obtaining
\begin{equation}
\begin{split}
\hstar (z) & = \left(\frac{\beta - 1}{\beta^2 - 1/4}\right)  \xhi (z) \lambdanu (z) \sigma_{0} f_{\rm esc} h\nu_0 \frac{dN_\gamma}{dM} \dot{\rho}_*(z) (1+z)^3 \\
& \approx  3 \times 10^{-19} \, {\rm erg \, s}^{-1}  \xhi (z) 
\left(\frac{f_{\rm esc}}{10^{-2}} \right) 
\\
& \times \left( \frac{\dot{\rho}_*(z)}{M_\odot \, {\rm Mpc}^{-3} \, {\rm yr}^{-1}} \right) (1+z)^{-2} \, .
\end{split}
\end{equation}

Finally, we choose the value for $f_{\rm esc}$ by assuming that the total optical depth, defined as
\begin{equation}\label{eq:tau_e}
\tau_e (z) = \int_0^{z} n_e (z') \sigma_T \, \frac{dl}{dz'} dz',
\end{equation}
does not exceed the $3\sigma$ value measured by PLANCK, $\tau_{e} = 0.066 \pm 0.016$ \citep{2016A&A...594A..13P}. 
In Fig.~\ref{fig:optdepth} we show the optical depth corresponding to $f_{\rm esc}=10^{-2}$ and compared to the observed value for $\tau_e$.  
We also show that our model predicts a fully reionized IGM by $z \sim 6$ as inferred from the Gunn-Peterson trough detections~\citep{2006AJ....131.1203F}.

Finally, we note that a very similar value for $f_{\rm esc}$ has been found by means of a more sophisticated approach and tested against more observables by the authors of~\citet{2005MNRAS.361..577C}.

\section{Cosmic rays in the IGM}\label{sec:cr}

CRs accelerated in early galaxies can act in principle as an additional source of non-thermal energy for the IGM.
In the Milky Way, most of the CR energy is in protons. They diffuse or advect out of the Galaxy on timescales of about 30 Myr that can be directly inferred from secondary-over-primary ratios~\citep{2013A&ARv..21...70B}, with only a few percent of the energy lost in pion production and ionization~\citep{2007ARNPS..57..285S}.
Thus, a large fraction of the power injected in CRs in our Galaxy ends up in the surrounding IGM.

In our model CRs are accelerated by star-forming galaxies with an universal energy spectrum and their energy released far beyond the circumgalactic gas, i.e. in the IGM. 
{ This conclusion is motivated by the fact that earlier structures are expected to be less confining than the present galaxies, since they were smaller and had a weaker magnetic field.}
{ In fact, \citet{2015MNRAS.448L..20L} and \citet{2006ApJ...651..658R} argued that primary CRs escape from parent galaxies on a timescale short enough so that they do not suffer any energy loss.} 

To follow the propagation of an homogeneous CR population in an expanding universe for a continuous source of CRs, we generalize the classical work of~\citet{1977ApJ...216..177M}, including all the relevant energy loss processes.

\subsection{Production in the early galaxies} \label{sec:production}

Star formation pumps energy into CR protons at a rate
\begin{equation}\label{eq:budget}
\dot{E}_{\rm p} (z) = \epsilon E_{\rm SN} {\rm SNR}(z) (1+z)^3,
\end{equation}
where $E_{\rm SN} \sim 10^{51}$~erg is the average explosion energy for a Type {\small II} supernova (SN) not going in neutrinos, $\epsilon \sim 0.1$ is the fraction of the kinetic energy transferred to CRs by a single SN, and SNR is the comoving SN rate. 
In principle, one should account also for Helium nuclei. For simplicity, we assume here that $\alpha$-particles can be treated as four protons, and hence be absorbed in the proton spectrum efficiency. 

In addition to the SFR, to derive SNR we need to know the number of SNe explosions per solar mass of forming stars, which is given by
\begin{equation}
f_{\rm SN} = \frac{\int_{8}^{50} \phi(m) dm}{\int_{0.1}^{100} m \phi(m) dm} \sim 10^{-2} \, M_\odot^{-1} ,
\end{equation}
where $\phi(m)$ is the Initial Mass Function (IMF) of Population II/I stars, for which we assume the following form
\begin{equation}
\phi(m) \propto m^{-1+x} \exp \left( -\frac{m_c}{m} \right) \, ,
\end{equation}
with $x=-1.35$, $m_c=0.35$~$M_\odot$; $m$ lies in the range $[0.1,100]$~$M_\odot$~\citep{1998MNRAS.301..569L}.
We neglect the contribution from Pop III stars. Their SF history is very debated and still highly uncertain. In fact, detailed studies, exploiting cosmological hydrodynamical simulations implementing chemical feedback effects, have shown that Pop II/I stars dominate the global SFR at any redshift~\citep{2007MNRAS.382..945T,2010MNRAS.407.1003M}. 
Moreover, the recent surveys hunting for PopIII stars in the Milky Way have found no metal free stars so far, implying that they are rare in the Milky Way even if they exist~\citep{2005ARA&A..43..531B}.

Combining the above formulae, Eq.~\eqref{eq:budget} becomes
\begin{multline}
\dot{E}_{\rm p} (z) \sim 10^{-33} \, {\rm erg} \, {\rm cm^{-3}} \, {\rm s^{-1}} \,
\left( \frac{\epsilon}{0.1} \right) 
\left( \frac{E_{\rm SN}}{10^{51} \, {\rm erg}} \right) \\
\left( \frac{f_{\rm SN}}{10^{-2}  M_\odot^{-1}} \right) 
\left( \frac{\dot{\rho}_* (z)}{M_\odot {\rm yr}^{-1} {\rm Mpc}^{-3}} \right) 
(1+z)^3  .   
\end{multline}

Particle acceleration in SN explosions is believed to occur through diffusive shock acceleration, which leads to momentum power-law spectra of accelerated particles.
With this in mind, we assume that the source function of volume averaged CR protons injected by SNe (defined as a rate per unit energy and volume) is 
\begin{equation} \label{eq:Q}
q_p(E,z) = 
%C(z) \frac{dN_p}{dE} = 
\frac{C(z)}{\beta(E)} \, \left( \frac{E^2 + 2 E m_p c^2}{E_0^2 + 2 E_0 m_p c^2} \right)^{-\frac{\alpha}{2}} \, , 
\end{equation}
where $E$ is the proton kinetic energy, $E_0 = 1$~GeV, $m_p$ is the proton mass, $\beta = v/c$ is the dimensionless velocity of the particle, $\alpha \geq 2$ is the slope of the differential spectrum of accelerated particles and $C(z)$ is a redshift-dependent normalization obtained by imposing that the total kinetic energy rate equals $\dot{E}_{\rm p} (z)$, i.e.
\begin{equation}\label{eq:energetic}
\dot{E}_{\rm p} (z) = \int_{E_{\rm min}}^{E_{\rm max}} E q_p(E,z) \, dE \, .
\end{equation}

In Eq.~\eqref{eq:energetic} we fix $E_{\rm max} = 10^6$~GeV \citep[][]{1983A&A...125..249L} and we verify a posteriori that our conclusions are not strongly dependent on our choice of $E_{\rm min}=10$ keV.
From Eq.~\eqref{eq:Q} one can easily realize that $\alpha$ is the key parameter determining the fraction of the total kinetic energy released that goes into protons with $E \ll 1$~GeV. We keep $\alpha$ as the only free parameter of the model.

\begin{figure}
\begin{center}
\includegraphics[width=.97\columnwidth]{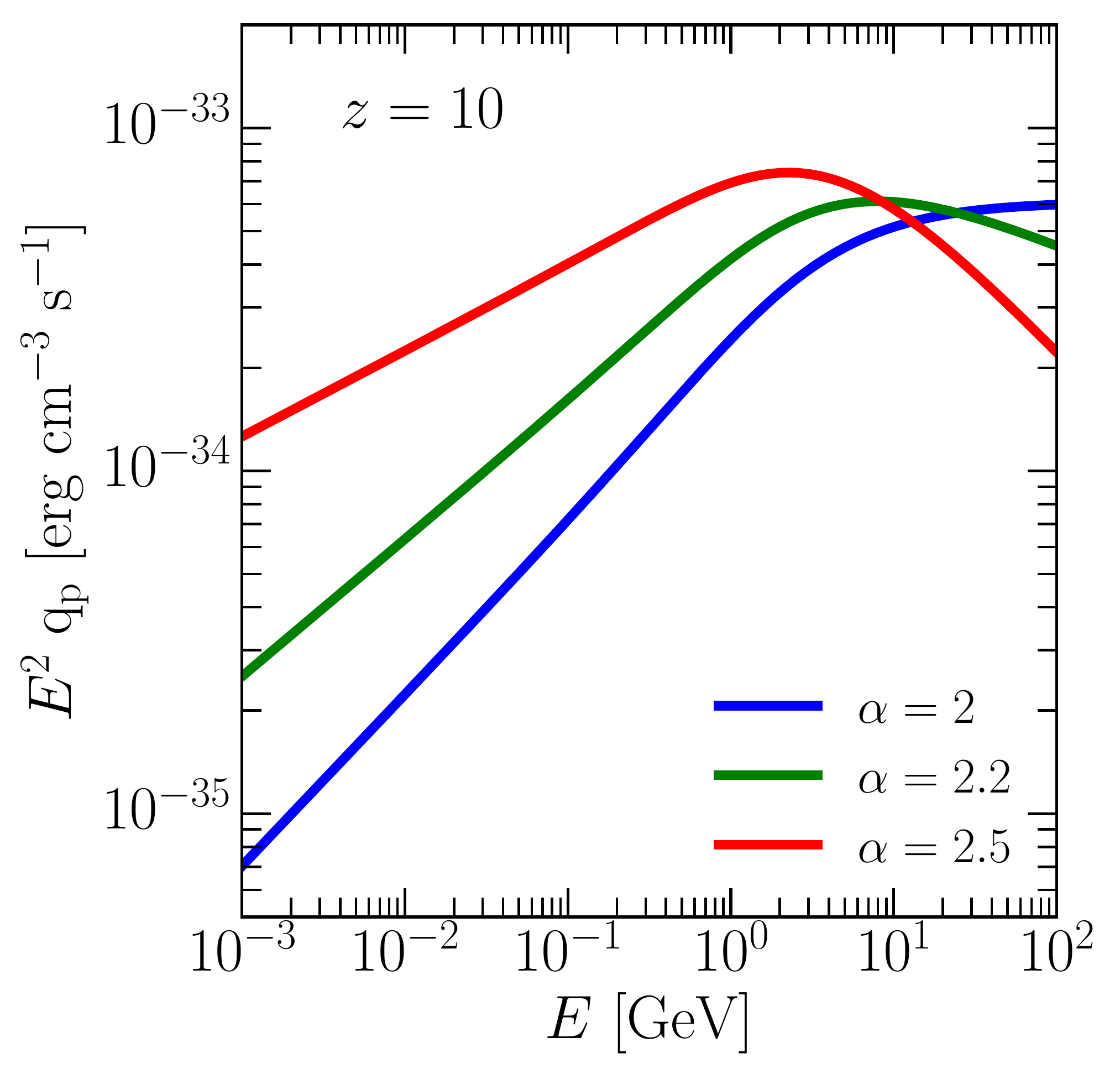}
\end{center}
\caption{Source function of CR protons with respect to their kinetic energy at $z=10$ for a spectrum slope $\alpha=2$ (blue line), 2.2 (green) and 2.5 (red).}
\label{fig:Q}
\end{figure}

In Fig.~\ref{fig:Q} we plot the source function as a function of the kinetic energy for three different values of $\alpha$.
We found that protons with kinetic energies below $10$~MeV represent $0.5$\%, $2.7$\%, $13$\% of the total kinetic energy released for $\alpha = 2, \, 2.2, \, 2.5$, respectively. 

\subsection{Energy losses in the IGM}\label{sec:elosses}

CRs can be an efficient heat source especially for a low density gas. When the CR proton ionizes an atom, it transfers a certain fraction of its kinetic energy to the electron, which is either used for further atomic excitation and ionization, or distributed via elastic collisions to other species of the medium. The latter process increases the kinetic temperature of the gas.

Ionization losses can be taken into account using the Bethe-Bloch equation, that for $\gamma\ll m_p/m_e$ can be approximated as~\citep[][]{2007A&A...473...41E}
\begin{equation}\label{eq:Ion}
\left. -\frac{dE}{dt} \right|_I = 
\frac{4\pi e^4}{m_e \beta c}\sum_Z Z x_{\rm HI} n_Z\left[\ln\left(\frac{2m_e c^2}{I_Z} P^2\right)-\beta^2\right] \, , 
\end{equation}
where $m_e$ is the electron mass, $n_Z$ is the number density of the elements with atomic number $Z$, $I_Z$ is the ionization potential ($I_{\rm H}=13.6$ eV and $I_{\rm He}=24.6$ eV), and $P = p / (m_p c^2) = \sqrt{\gamma^2 - 1}$ is the dimensionless particle momentum.

Losses due to Coulomb interactions, which describe the fact that the energy lost by protons in Coulomb interactions is directly transferred to momentum of the plasma electrons (and hence heating), can be expressed as~\citep{1972Phy....62..555G}
\begin{equation} \label{eq:Cou}
\left. -\frac{dE}{dt} \right|_C = 
\frac{4\pi e^4 n_e}{m_e \beta c}\left[\ln\left(\frac{2m_e c^2\beta}{\hbar \omega_{pl}} P \right)-\frac{\beta^2}{2}\right] \, , 
\end{equation}
where $\omega_{pl}= (4\pi e^2 n_e/m_e)^{1/2}$ is the plasma frequency.
The number density of free electrons $n_e$ is computed from our reionization model described in~\S\ref{sec:reion} assuming only stellar radiation ionizations. 
Indeed, we verify a posteriori that CR ionizations are a subdominant contribution~(see \S\ref{sec:reionization}).

Inverse Compton scattering with respect to CMB photons can be safely neglected, since its timescale is much longer compared to collisional processes~\citep{2012MNRAS.422..420E}. 

Finally, the adiabatic energy losses caused by Hubble expansion can be taken into account as~\citep{1977ApJ...216..177M}
\begin{equation} \label{eq:Adi}
\left. -\frac{dE}{dt} \right|_a = \frac{E (E+2 m_p c^2)}{E+ m_p c^2} \frac{1}{1+z}\frac{dz}{dt} \, .
\end{equation}

For CR energy deposition to be effective, its timescale must be shorter than the Hubble time,
\begin{equation}
t_i = \frac{E}{dE_i/dt} \leq t_H \, ,
\end{equation}
where
\begin{equation}\label{Eq:thubble}
t_H(z)  = \int_\infty^z dz' \, \frac{dt}{dz'} \simeq \frac{2(1+z)^{-3/2}}{3H_0\Omega_{m}^{1/2}} \simeq 0.2 \, \left(\frac{1+z}{21} \right)^{-3/2} {\rm Gyr}.
\end{equation}

\begin{figure}
\begin{center}
\includegraphics[width=0.97\columnwidth]{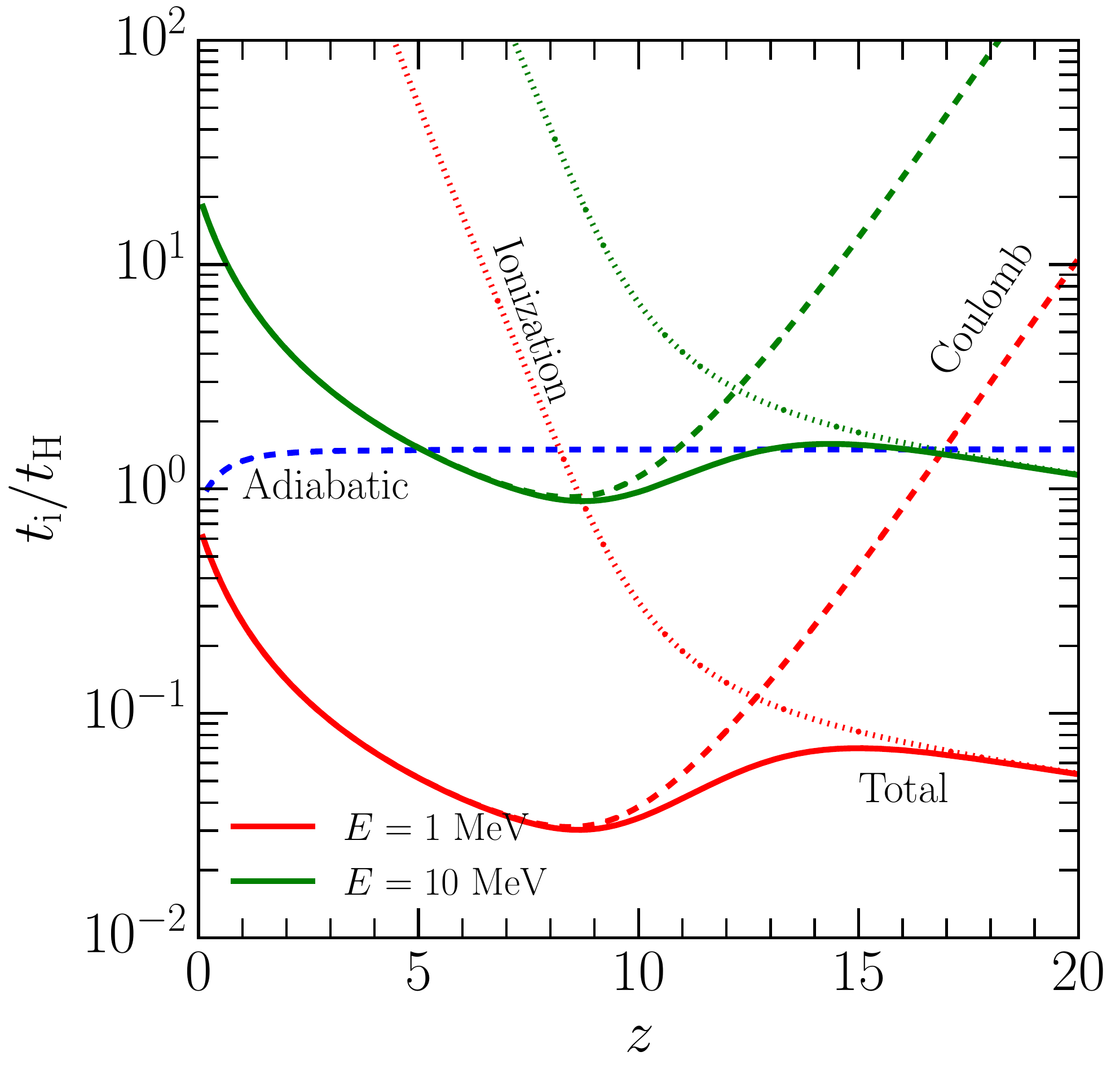}
\caption{Energy loss timescales (see Eqs.~\eqref{eq:Ion}, \eqref{eq:Cou} and \eqref{eq:Adi}) normalized to the Hubble time for CR protons of 1 and 10 MeV. { The adiabatic time scale (blue dashed line) is independent from thefrom the particle energy}.}\label{fig:eloss}
\end{center}
\end{figure}

In Fig.~\ref{fig:eloss} we plot $t_i/ t_H$ as a function of redshift for the loss mechanisms discussed above. 
Energy losses are efficient for kinetic energies $\lesssim 10$ MeV. Ionization losses dominate over Coulomb losses at earlier epochs when the IGM was mainly neutral.

\subsection{Propagation in the IGM}

The evolution equation of the CR proton number density (averaged over the volume), $n_p(E,z)$, can be written as follows~\citep{1977ApJ...216..177M,2006ApJ...651..658R,2008MNRAS.390L..14E}
\begin{equation} \label{eq:transp}
\frac{\partial}{\partial t}  N_p + \frac{\partial}{\partial E} \left( b N_p \right) + \frac{N_p}{t_D} = Q_p(E,z) \, ,
\end{equation}
where the number density of protons and the source term are now normalized to $n_{\rm H}(z)$, being $N_p = n_p / n_H$ and $Q_p = q_p / n_H$. 
We also assume that the total energy loss rate $b \equiv dE/dt$ is given by the sum of the loss processes described by Eqs.~\eqref{eq:Ion}, \eqref{eq:Cou} and \eqref{eq:Adi}.

Proton-proton interactions cause CR energy loss in a timescale %that can be estimated using
\citep{2007Ap&SS.309..365G}
\begin{equation}
t_{\rm pp}^{-1} = n_{b} c \kappa \sigma_{\rm pp} \, , 
\end{equation}
where the inelasticity coefficient is $\kappa \approx 0.45$ and $\sigma_{\rm pp} \approx 35$~mb is the total inelastic cross-section for proton-proton interaction.
This results in $t_{\rm pp} \approx 10^{9}(1+z)^3$~Gyr, making this process negligible for our purposes. 

\begin{figure}
\begin{center}
\includegraphics[width=0.98\columnwidth]{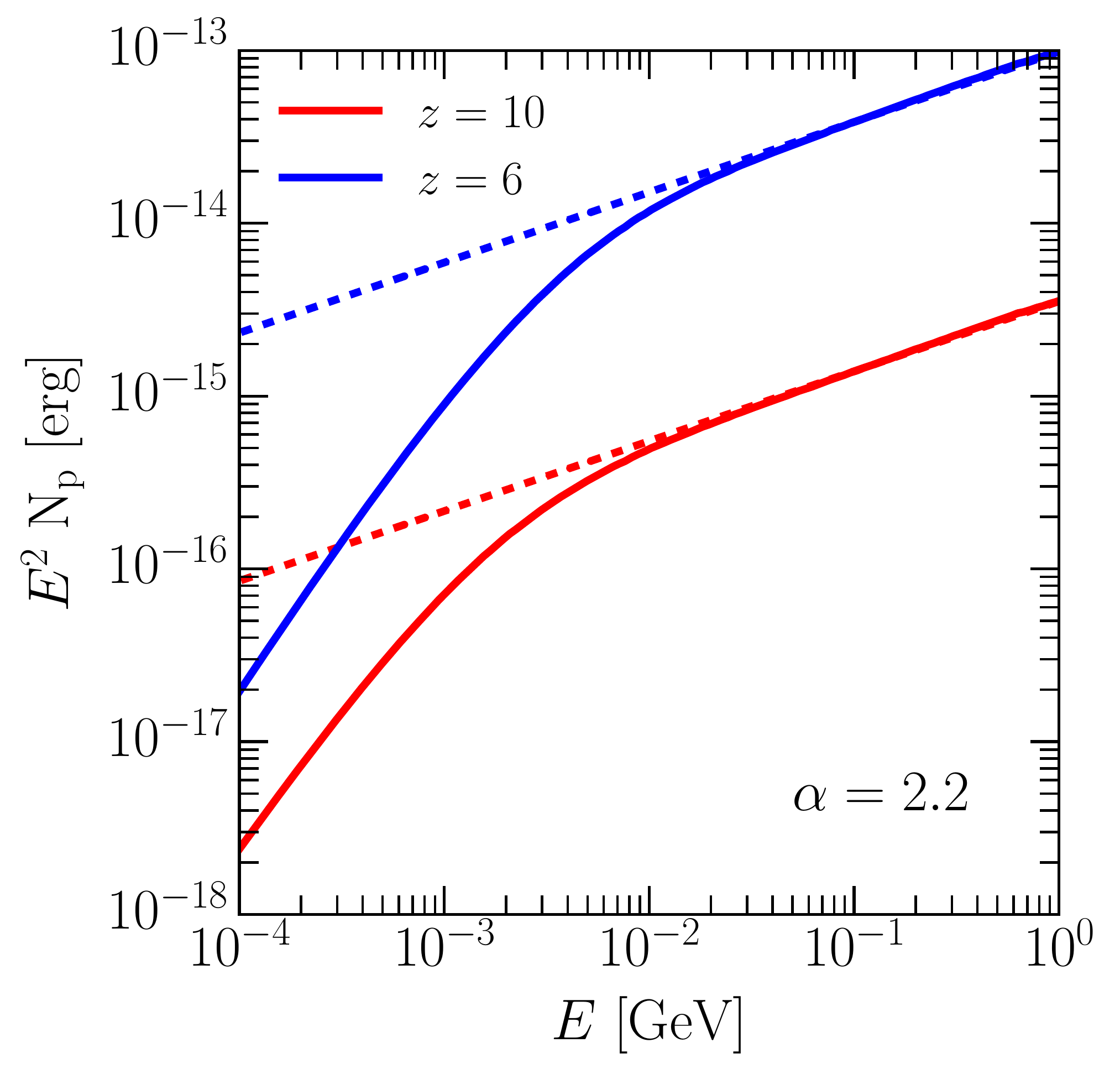}
\caption{Normalized CR proton number, $N_p$, as a function of energy at different redshifts. 
Solid (dashed) lines represent the solution of Eq.~\eqref{eq:transp} with (without) inclusion of the energy loss term. Blue and red lines refer to $z=6$ and 10, respectively.}\label{fig:N_p} 
\end{center}
\end{figure}

Eq.~\eqref{eq:transp} is solved numerically using the Crank-Nicolson implicit method described in~\citet{1992nrfa.book.....P}. The results are presented in Fig.~\ref{fig:N_p}, showing the redshift evolution of the proton spectrum. The effect of energy losses (solid vs. dashed lines) is evident mainly at low-energies ($E \lesssim 10$~MeV). 

\subsection{IGM ionization and heating}
We come now to the central question: can CR energy losses sensibly affect the IGM ionization state and/or temperature? 
The primary ionization rate for H is 
\begin{equation} \label{eq:ion_CR}
\Gamma_{\rm ion}^{\rm CR} = \frac{1}{W_H} \int_{E_{\rm min}}^\infty \left| \frac{dE}{dt} \right|_{\rm I} n_p(E) dE ,
\end{equation}
where $W_{\rm H}\simeq 36.3$~eV is the mean energy expended by a CR proton to create an ion pair~\citep{2007MNRAS.380..417J}.

Following~\citet{1969ApJ...158..161S}, we account for all secondary and higher generation ionizations by multiplying the primary ionization rate in Eq.~\eqref{eq:ion_CR} by a factor $\xi(x_e)$. 
For this we use a linear interpolation between the extreme case $\xi(1) = 1$ and  $\xi(0) = 5/3$, which reads as
\begin{equation}
%\xi(x_e) = \frac{5}{3} (1 - x_e)
\xi(x_e) = \frac{5}{3} - \frac{2}{3} x_e .
\end{equation}

In the case of Coulomb losses, we can assume that all the lost energy is entirely converted into background heat. 
The corresponding heating rate can be calculated by using Eq.~\eqref{eq:Cou} 
\begin{equation}
\mathcal{H}_{\rm C} =  \int_{E_{\rm min}}^\infty \left| \frac{dE}{dt} \right|_{\rm C} n_p(E) dE \, .
\end{equation}

The contribution to heating by secondary electrons from ionization can be divided in three regimes according to their energy: for $E > I_{\rm H}$ ionization or excitation of H {\small I} can occur; for $3 \, I_{\rm H} / 4 < E < I_{\rm H}$ the electron can suffer losses by Coulomb and excitation collisions; for $E < 3 \, I_{\rm H} / 4$ the energy is transferred directly into heating. 

An approximate general formula for the heating is given by~\citet{2007MNRAS.380..417J}, leading to a total heating rate by CR as
\begin{equation} \label{eq:heat_CR}
\mathcal{H}^{\rm CR} = \left[W_{\rm H} - \xi(x_{\rm e}) I_{\rm H} \right] \Gamma_{\rm ion}^{\rm CR} + \mathcal{H}_{\rm C} \, .
% + \left(E-\frac{3}{4}I_Z\right) \Lambda_{\rm exc},
\end{equation}
It follows that, in a neutral medium, a heat input of $\Delta E = W_{\rm H} - 5/3 \, I_{\rm H} \sim 13.6$~eV for every ionization of hydrogen via CR protons is transferred  to the IGM. 
We note, however, that the above expression is likely to overestimate the heating rate as electron energy losses via excitations are not accounted for. 

\section{Results}\label{sec:results}

\begin{figure}
\begin{center}
\includegraphics[width=.97\columnwidth]{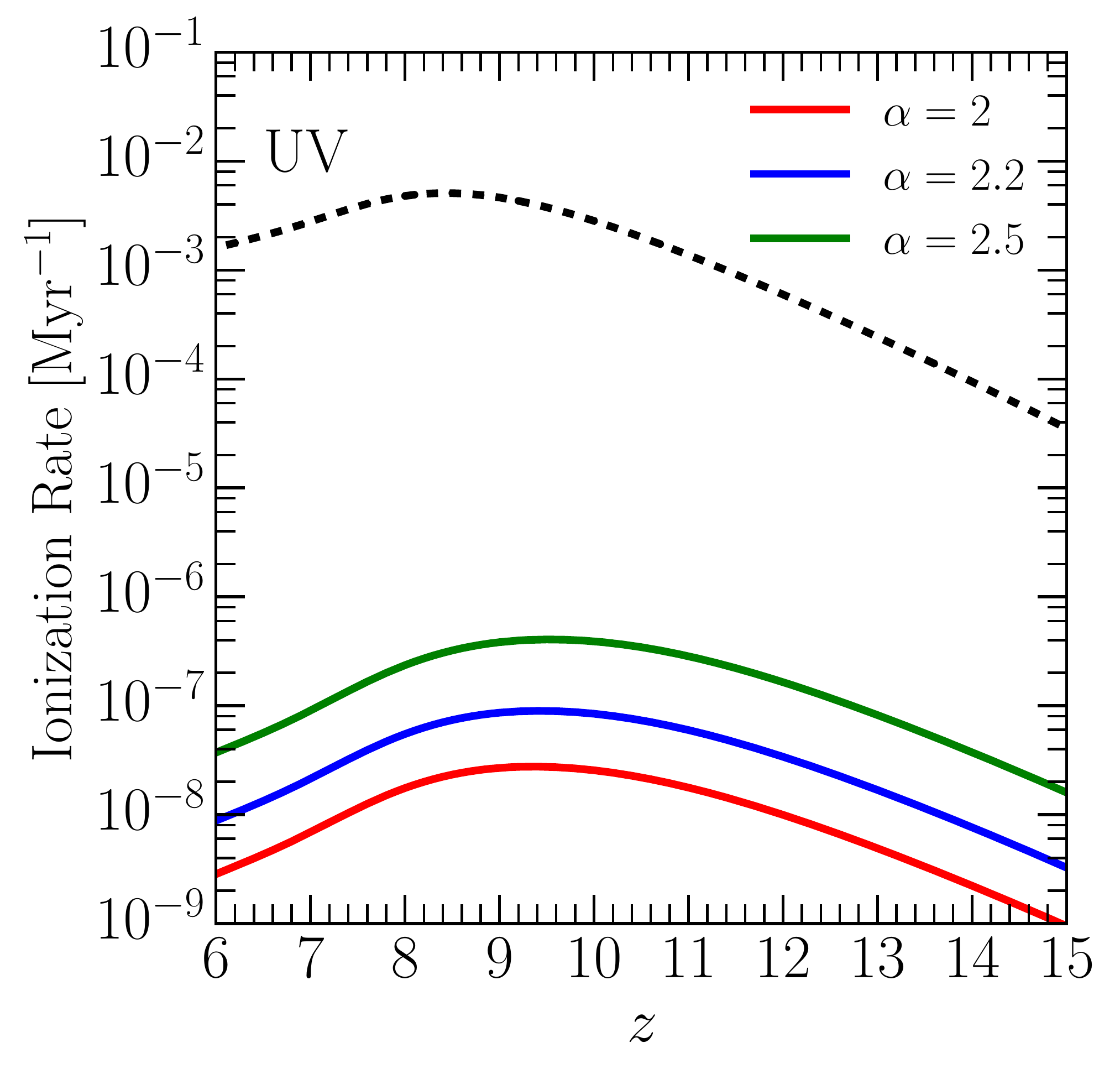}
\caption{The ionization rate as a function of $z$ from CR (solid lines) for different injection slopes and compared with the ionization rate from stellar UV photons $I_{\rm UV} = x_{\rm HI} \Gamma_{\rm HI}$ (dashed line).}
\label{fig:ion_CR}
\end{center}
\end{figure}

We are now ready to discuss the effects of CRs on the IGM ionization fraction and temperature from our model.

\subsection{Impact of CRs on reionization} \label{sec:reionization}

The ionization rate computed with Eq.~\eqref{eq:ion_CR} is shown in Fig.~\ref{fig:ion_CR} and compared to the ionization rate by UV photons through Eq.~\eqref{eq:ion_rate}. 
The CR ionization rate is several orders of magnitude smaller than the UV photoionization rate.
This justifies the fact that CR were not included when we described our reionization model in \S~\ref{sec:reion}.
This result can be better understood by comparing the stellar and CR emissivities in a more simplified scenario.
We recall that the ionizing photon emissivity by galaxies is given by
\begin{equation}
\epsilon_* = f_{\rm esc} E_\gamma f_{\rm SN} \dot{\rho}_* \, ,
\end{equation}
where the energy in photons is given by 
\begin{equation}
E_\gamma = \dot{N}_\gamma h \nu_0 t_* \, ,
\end{equation}
where $\dot{N}_\gamma \sim 5 \times 10^{47}$~s$^{-1}$ is the rate of ionizing photons, and $t_* \sim 30$~Myr is the stellar lifetime. 
In doing so, we are assuming that the UV emission by galaxies is dominated by the same stars that go supernova, with mass $\sim 10$~M$_\odot$.

On the other hand, the CR emissivity can be written as
\begin{equation}
\epsilon_{\rm CR} = f_d \epsilon E_{\rm SN} f_{\rm SN} \dot{\rho}_* \, ,
\end{equation}
where $f_d \sim 10^{-3}$ is the fraction of energy deposited in the IGM and, as we discussed in \S~\ref{sec:elosses},  corresponds to the fraction of energy in CR protons with $E \lesssim 10$~MeV. We are thus assuming that all the deposited energy is used to ionize IGM atoms.

The ratio between the corresponding fluxes is then
\begin{equation}
\frac{J_{\rm CR}}{J_*} = \frac{\epsilon_{\rm CR} \lambda_{\rm CR}}{\epsilon_* \lambda_*} \sim 10^{-3} \frac{\lambda_{\rm CR}}{\lambda_*} \, ,
\end{equation}
where $\lambda_i$ designates the corresponding mean free path. 

Finally, the ratio of the ionization rates, $J_i/\tau_i$, with $\tau_i$ the corresponding energy loss time, can then be roughly estimated as $10^{-3}$ by approximating $\tau_i \propto \lambda_i$.

\begin{figure}
\begin{center}
\includegraphics[width=.97\columnwidth]{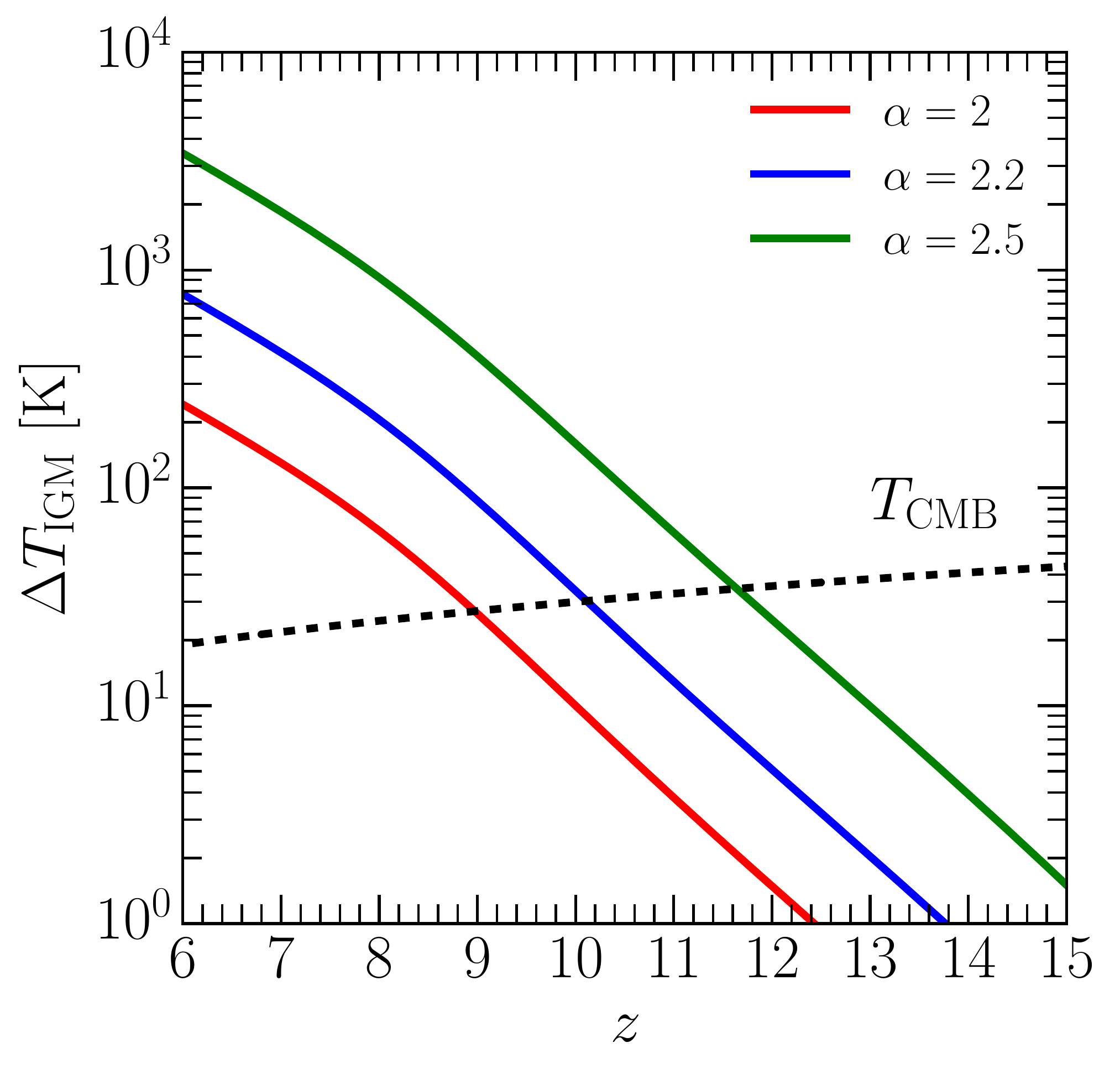}
\caption{Increment of the average IGM temperature by CRs as a function of redshift for three values of the CR injection slope. The CMB temperature at the same redshift is shown by the dashed line.}
\label{fig:heat}
\end{center}
\end{figure}

\subsection{IGM heating}

While there is a consensus that UV stellar radiation is largely responsible for cosmic reionization, its impact on the global IGM temperature is limited to fully ionized regions.
For quasi-neutral IGM regions, X-rays are clearly more relevant due to their larger mean free path. 
These energetic photons might come from an early population of relatively soft X-ray binaries ~\citep{2013MNRAS.431..621M}.
Alternatively, if sourced by black holes, they could pre-heat the gas up to  $10^4$ K ~\citep{2004MNRAS.352..547R}.    
However, uncertainties related to the nature and abundance of their sources at high redshift make predictions very uncertain~\citep{2012MNRAS.426.1349M, 2007MNRAS.376.1680P}.
For this reason, CRs might represent a competitive, alternative source of thermal input for the neutral IGM.

The IGM temperature increase produced by CR heating (Fig.~\ref{fig:heat}) is 
\begin{equation}
\Delta T (z) = \frac{2}{3} \frac{\mathcal{H}^{\rm CR}(z)}{k_B H(z)} ,
\end{equation}
where $\mathcal{H}_{\rm h}^{\rm CR}$ is given by Eq.~\eqref{eq:heat_CR}. The IGM temperature can be raised up to $\sim 3 \times 10^3$~K before reionization is complete and it exceeds the CMB one, $T_{\rm CMB}(z) = 2.725 \, (1+z)$~K, at $z \lesssim 9 (12)$ for $\alpha = 2(2.5)$. These results imply that the IGM is pre-heated well before being reionized. 
Depending on the efficiency of the CR diffusion mechanism, pre-heating might be confined in regions around star forming galaxies, or, if diffusion is very efficient, it might give rise to a more distributed, quasi-uniform warmer floor (see \S~\ref{sec:diffusion}).

We can directly compare our predictions with the results presented in Fig.~1 of~\cite{2015MNRAS.454.3464S}. These authors also study the IGM heating from low-energy CRs, finding a temperature increament between (1-10$^3$) K, mostly depending on the minimum halo mass allowed to form stars.
The highest temperature was found for a supernova energy explosion $E_{\rm SN}=10^{53}$ erg (hence corresponding to a PopIII supernova) out of which 5\% is pumped into low-energy CRs, and a minimum star-forming halo mass of $3\times 10^5\, M_\odot$. 
Our model predicts a similar $\Delta T$ but relying on fairly standard stellar populations (and supernova explosion energies) whose SFR has been calibrated with both the observed cosmic star formation history and reionization constraints from the Thomson scattering optical depth.

\subsection{Diffusion in the IGM}\label{sec:diffusion}

As energetic particles are injected into the IGM, they will be subject to a random-walk through it. 
The travelled distance depends on the strength and structure of the intergalactic magnetic field, on which very little is known at high redshift. The standard assumption is to consider CR energy deposition as a uniform background~\citep[see, e.g.,][]{2015MNRAS.448L..20L}. 
The rationale behind this assumption is based on the following argument.  

The slowest rate at which CRs can diffuse corresponds to the so-called Bohm diffusion. This assumes one scattering per gyroradius; the diffusion coefficient is $D_B = c r_L / 3$, where $r_L$ is the particle Larmor radius.
The \emph{maximum} diffusion timescale between haloes is then given by
\begin{equation} \label{eq:t_B}
t_B \simeq \frac{\langle d \rangle^2}{D_B},
\end{equation}
where $\langle d \rangle$ is the average {\it proper} distance between them, 
and the Bohm diffusion coefficient for protons ($Z=1$) can be estimated as 
\begin{equation} \label{eq:bohmd}
\begin{split}
D_B (p,z) \sim 1.1 \, \frac{{\rm Mpc}^2}{\rm Gyr} \, \left( \frac{p}{\rm GeV/c} \right) \left( \frac{B_0}{10^{-16} {\rm G}} \right)^{-1} \left(\frac{1+z}{21}\right)^{-2} \, ,
\end{split}
\end{equation}
$B_0=10^{-16}$ G is the assumed IGM magnetic field strength at $z=20$, following~\citet{2015MNRAS.454.3464S}.

To estimate $\langle d \rangle$, we consider uniformly distributed haloes. Their average inter-distance is then 
\begin{equation} \label{eq:d_h}
\frac{4\pi}{3} \langle d \rangle^3 \sim \left( M_h \frac{dN}{dM_h} \right)^{-1} (1+z)^{-3} \, ,
%\sim 3 \times 10^{-2} \left( \frac{M_h \, dN/dM_h}{\rm Mpc^{-3}} \right)^{-\frac{1}{3}} \left( \frac{1+z}{21} \right)^{-1} \, {\rm Mpc} 
\end{equation}
where $dN/dM_h$ is the comoving halo density. At $z=20$, the haloes contributing mostly to the SFR are those with $M_h \sim M_{\rm min}(z = 20)$, whose mean separation is $\langle d \rangle \sim 50$~kpc. 
From Eq.~\eqref{eq:t_B} and Eq.~\eqref{Eq:thubble}, we deduce that $t_B \lesssim t_H$ as long as the CR energy is $E \gtrsim 20$~keV. This result would support the idea of a uniform thermal deposition by CRs.

However, the above calculation is incomplete, as CRs may affect the environment in which they propagate. When CRs escape from the halo they produce an electric current to which the background plasma reacts by generating a return current that in turn leads to the development of small scale instabilities. The growth of such instabilities leads to large turbulent magnetic fields and to an enhanced particle scattering. In short, CRs may undergo self-confinement~\citep{Blasi:2015esa}.
In this scenario, the particle diffusion timescale can be significantly larger than what we found in Eq.~\eqref{eq:t_B}.

In order to get an estimate of the potential maximum effect associated with this mechanism, we generalize the formalism developed by~\citet{Blasi:2015esa} to the non-relativistic regime; moreover, we maximize the effect by assuming that all escaping CRs contribute to the self-generated magnetic field.

The differential number density (in momentum) of CRs escaping out from a halo at a distance $r$ from it can be written as\footnote{The injection spectrum assumed here is equivalent to Eq.~\eqref{eq:Q}, since $dN/d^3p \propto p^{-4}$ corresponds to $dN/dE \propto p^{-2}$.}
\begin{equation}
f(p,r) = \frac{dN_{\rm p}}{dV d^3p} = A(r) \left( \frac{p}{p_0} \right)^{-4} \, ,
\label{eq:inj_spec}
\end{equation}
where $A(r)$ is obtained by imposing that the total pressure exerted on a surface $S = 4 \pi r^2$ by the CR source,
\begin{equation}\label{eq:ps}
P_{\rm s} = \frac{F_{\rm s}}{S} \sim \frac{L_{\rm CR}/c}{4 \pi r^2} \, ,
\end{equation}
is given by the CR pressure at the same distance
\begin{equation}\label{eq:pcr}
P_{\rm CR} \sim \int_{p_{\rm min}}^{p_{\rm max}} dp \, p^3 v(p) f(p,r) \, ,
\end{equation}
where the source luminosity in CRs, $L_{\rm CR}$, for a typical star forming halo at $z=20$ is
\begin{equation}
L_{\rm CR} = f_* f_{\rm SN} E_{\rm CR} \frac{\Omega_b}{\Omega_m}\frac{M_h}{t_{\rm ff}} \sim 10^{38} \, \text{erg s}^{-1}. %\sim 6.2 \times 10^{40} \, \text{GeV/s}
\end{equation}
By equating Eqs.~\eqref{eq:ps} and \eqref{eq:pcr}, one has
\begin{equation}
A(r) = \frac{L_{\rm CR}}{4 \pi r^2 c} \left[ \int_{p_{\rm min}}^{p_{\rm max}} dp \, p^3 v(p) \left( \frac{p}{p_0} \right)^{-4} \right]^{-1} \, .
\end{equation}
The electric current $j_{\rm CR}$ associated with CRs streaming away from their sources can be written as 
\begin{equation}\label{eq:current}
j_{CR}(p) = e \int_p^{p_{\rm max}} 4 \pi p^2 dp \, v(p) f(p,r) \sim \frac{e L_{\rm CR}}{c r^2 p_0} g(p/p_0) \, ,
\end{equation}
where we have introduced 
\begin{equation}
g(x) = \frac{\int_{x}^{x_{\rm max}} dx \beta(x p_0) x^{-2}}{\int_{x_{\rm min}}^{x_{\rm max}} dx \, \beta(x p_0) x^{-1}} \, .
\end{equation}

Assuming that the non-resonant modes are able to grow on a timescale much shorter than $t_H$, the magnetic field saturates at a value $\delta B_{\rm s}$. The saturation level is set by equipartition between the energy density of the amplified field and the kinetic energy density of the CR current (see Eq.~\eqref{eq:current}):
\begin{equation}\label{eq:B_sg}
\frac{\delta B_{\rm s}^2}{8 \pi} \sim \frac{L_{\rm CR}}{c r^2} \left[ \frac{p}{p_0} g(p) \right]_{\rm max} \sim 0.01 \frac{L_{\rm CR}}{c r^2} \, ; %= \sqrt{\frac{p}{c} g(p)} \, \, \frac{L_{CR}^{1/2}}{r}
\end{equation}
the last approximate equality holds since $p g(p)$ reaches a maximum at $\sim$GeV/c and remains constant for larger momenta. Numerically this yields $\delta B_{\rm s} \approx 0.01 \mu$G at $r=1$ kpc.

The corresponding mean free path at a given epoch can be finally computed assuming Bohm diffusion, as in Eq.~\eqref{eq:bohmd}, $\lambda_{\rm CR} = \sqrt{t_B D_B}$. The magnetic field is given by Eq.~\eqref{eq:B_sg} with $r$ representing now the mean free path. This results in
\begin{equation}
\lambda_{\rm CR} = 1 \, \text{kpc} \, \left( \frac{t_{i}}{\text{Gyr}} \right)\left( \frac{L_{\rm CR}}{10^{38} \, \text{erg s}^{-1}} \right)^{-1/2} \left( \frac{p}{\rm GeV/c} \right) \, .
\end{equation}

If this is the case, it would imply that CRs heating is far from uniform; rather, it is highly patchy and clustered around the smallest star forming halos.

In practice, a number of neglected effects might reduce the efficiency of CR self-confinement. These are: 
(a) the presence of neutrals outside the fully ionized bubbles can quickly damp the waves generated through the CR streaming instability;
(b) the B-field equipartition value in Eq.~\eqref{eq:B_sg} might not be attained due to an inefficient CR-magnetic energy density conversion. 
Moreover, a sufficiently strong intergalactic B-field and/or a smaller galactic CR luminosity may result in a magnetic-CR energy density ratio that is too large for the development of the instability in the non-resonant regime.

We note that observations of the redshifted HI 21 cm line from these high redshifts would be very sensitive to the morphology of the pre-heated, neutral regions. We thus expect that the clustered heating scenario leaves unique imprints in the power spectrum of such  radiation. Additionally, the analysis of the power spectrum should allow to discriminate between the case in which the heating source are X-rays or CRs. Finally, we could also gain precious information about the strength and structure of early intergalactic magnetic fields and the efficiency of CR acceleration by the first SNe. All these aspects are very hard to investigate with any other mean.

\section{Conclusions}\label{sec:conclusions}

In this work we have shown that CRs can influence the temperature and ionization fraction of the IGM using a self-consistent model for galaxy formation and cosmic reionization. The model was designed to reproduce the observed SFR at redshift $z \lesssim 10$, and a cosmic history consistent the latest PLANCK results. 
Such data constrain the conversion efficiency of gas into stars ($f_* = 0.04$) and the population-averaged escape fraction of ionizing photons into the IGM ($f_{\rm esc} = 0.01$). From the supernova rate evolution given by the model we further derived the CR energy density. 

CRs with energies $ < 1 $ MeV lose energy predominantly by ionizations at redshifts $z > 10$, and by Coulomb scatterings at lower redshifts (see Fig.~\ref{fig:eloss}). 
The energy lost via Coulomb collisions goes directly to heat, increasing the IGM temperature at $z \sim 10$ above the standard adiabatic thermal evolution $\Delta T \sim 10-200$~K, depending on  the slope of the CR injection spectrum in the range $2 < \alpha < 2.5$, and on the transport efficiency of CRs out of the first star-forming structures. 
Such increase is comparable or higher than that produced by two other popular heating mechanisms, i.e. X-rays~\citep{2013MNRAS.431..621M} and dark matter annihilations~\citep{2014JCAP...11..024E}.
Plasma instabilities induced by blazar TeV photons can additionally heat up the IGM above the temperature induced by photo-heating~\citep{2012ApJ...752...23C,2016ApJ...833..118C}. Compared to our results, this mechanism is relevant at lower redshifts, $z\lesssim 6$, and provides a more uniform background. The signal yielded by such contribution would therefore be easily distinguishable from the one expected from CRs. 

Our model for the CR injection and transport in the IGM is based on two commonly accepted assumptions: (1) CRs escape from star-forming structures on a timescale much shorter than the energy loss timescale in the ISM, (2) CRs provide a spatially uniform energy density floor, i.e. a ``background''.

The first assumption is certainly valid for Milky Way protons with $E > 100$~MeV. Whether it holds for high-$z$ galaxies depends on poorly known quantities, such as the turbulent magnetic fields in these objects. On general grounds, however, weaker magnetic fields should correspond to a larger diffusion coefficient and a smaller size of the magnetic halo. Both factors lead to a shorter diffusion timescale than in the Galaxy.

The second assumption has been carefully investigated in \S~\ref{sec:diffusion}.
We showed that CRs escaping from galaxies trigger streaming instabilities eventually amplifying the seed magnetic field up to equipartition.
Under the most optimistic conditions for the development of the instabilities, such self-generated magnetic field might efficiently
confine GeV particles around haloes for a time largely exceeding the Hubble time at $z \sim 20$. 
If true, this strongly clustered emission is expected to leave a specific imprint on the 21cm line power spectrum. Such detection would allow for the first time to study the structure and strength of magnetic fields in the Dark Ages.
However, a testable prediction of this CR heating signature requires a more detailed model for the ejection and propagation of CRs in the pre-ionized bubbles and will be investigated in a forthcoming work.

\section*{Acknowledgments}

N.L.~thanks GSSI in L'Aquila for the warm hospitality during the preparation of the paper. { We thank A.~Mesinger for useful discussions.} This work was partially supported by the ''Helmholtz Alliance for Astroparticle Physics (HAP)'' funded by the Initiative and Networking Fund of the Helmholtz Association, and by the Deutsche Forschungsgemeinschaft (DFG) through the Collaborative Research Centre SFB 676 ''Particles, Strings and the Early Universe''.

\bibliography{cr_igm_2016}
\bibliographystyle{mn2e}

\label{lastpage}

\end{document}